\documentclass{elsart}

\usepackage{graphicx,xspace}
\usepackage{epsfig}
\usepackage{amsmath,amssymb}
\usepackage{dcolumn}
\usepackage{bm}

\newcommand{\diffunit}{cm$^{-2}$s$^{-1}$sr$^{-1}$GeV}
\newcommand{\pointunit}{cm$^{-2}$s$^{-1}$GeV}
\newcommand{\dNdE}{E^{2}_{\nu} \times dN_{\nu}/dE_{\nu}}
\newcommand{\Nch}{$N_{\mathrm{ch}}$}
\newcommand{\ea}{{\it et al.}}
\newcommand{\ic}{IceCube}

\begin{document}

\begin{frontmatter}

\title{Sensitivity of the IceCube Detector to Astrophysical
Sources of High Energy Muon Neutrinos}
{\tiny 
\author[15]{J.~Ahrens}, 
\author[8]{J.N.~Bahcall}, 
\author[1]{X.~Bai}, 
\author[14]{R.C.~Bay}, 
\author[15]{T.~Becka}, 
\author[2]{K.-H.~Becker}, 
\author[16]{D.~Berley}, 
\author[6]{E.~Bernardini}, 
\author[3]{D.~Bertrand}, 
\author[9]{D.Z.~Besson}, 
% ...

\address[1]{Bartol Research Institute, University of Delaware, Newark, DE 19716, USA}
\address[2]{Fachbereich 8 Physik, BUGH Wuppertal, D-42097 Wuppertal, Germany}
\address[3]{Universit\'e Libre de Bruxelles, Science Faculty CP230, Boulevard du Triomphe, B-1050 Brussels, Belgium}
% ...
\address[6]{DESY-Zeuthen, D-15738 Zeuthen, Germany}
% ...
\address[8]{Institute for Advanced Study, Princeton, NJ 08540, USA}
\address[9]{Dept. of Physics and Astronomy, University of Kansas, Lawrence, KS 66045, USA}
% ...
\address[14]{Dept. of Physics, University of California, Berkeley, CA 94720, USA}
\address[15]{Institute of Physics, University of Mainz, Staudinger Weg 7, D-55099 Mainz, Germany}
\address[16]{Dept. of Physics, University of Maryland, College Park, MD 20742, USA}
% ...

}

\clearpage
\newpage

{\tiny
\begin{abstract}
We present the results of a Monte-Carlo study of the sensitivity of
the planned IceCube detector to predicted fluxes of muon neutrinos at
 TeV to PeV energies. A complete simulation of the 
detector and data analysis  is used to study the detector's capability
to search for muon neutrinos from sources such as active galaxies and gamma-ray
bursts. We study the effective  area and the angular resolution of the
 detector as a function of muon energy and angle of incidence.  
 We present detailed calculations of the sensitivity of the detector to
both diffuse and pointlike neutrino emissions, including an assessment of the
sensitivity to neutrinos detected in coincidence with gamma-ray burst 
observations.
After three years of datataking,  IceCube will have been able to detect
a point source flux of $E_{\nu}^2 \times dN_{\nu}/dE_{\nu} = 7 \times 10^{-9}$\,\pointunit~
at a 5-sigma significance, or, in the absence of a signal,
 place a 90\% c.l. limit at a level
$E_{\nu}^2 \times dN_{\nu}/dE_{\nu} = 2 \times 10^{-9}$\,\pointunit. 
A diffuse $E^{-2}$ flux  would be detectable at a minimum strength of
$ E_{\nu}^2 \times dN_{\nu}/dE_{\nu} = 1 \times 10^{-8}$\,\diffunit.
A gamma-ray
burst model following the formulation of Waxman and Bahcall 
would result in a 5-sigma effect after the observation of 200 bursts 
in coincidence with satellite observations of the gamma-rays.  
\end{abstract}
}


\end{frontmatter}

%\clearpage
\section{Introduction}
\label{sec:intro}
The emerging field of high-energy neutrino astronomy 
\cite{GHS95,L&M,CS} has seen
the construction, operation and results from the first detectors, and
proposals for the next generation of such instruments. The first pioneering
efforts of the DUMAND \cite{dumand} collaboration were followed by the 
successful
deployments of NT-200 at Lake Baikal \cite{baikal1} and AMANDA \cite{amanda1} 
at the South
Pole. These detectors have demonstrated the feasibility
of large neutrino telescopes in open media like water or ice. They
have observed neutrinos produced in the atmosphere \cite{atmospheric} and have
put limits on the flux of extraterrestrial neutrinos  \cite{baikal2,diffuse},
which are significantly
below those obtained from the much smaller underground neutrino detectors 
\cite{macro2,superk}. 
The results obtained so far, and further refinements to astrophysical
theories of extra-terrestrial neutrino fluxes from cosmic sources 
have provided the impetus to construct
a neutrino observatory on a much larger scale. 
Proposals for a detector in the deep water of the Mediterranean have come 
from the 
ANTARES \cite{antares}, NESTOR \cite{nestor} and NEMO \cite{nemo} 
collaborations. 
IceCube is a projected cubic kilometer under ice neutrino detector 
\cite{icpropusa,icpropger,PDD}, to be located near the geographic 
South Pole in Antarctica.

The IceCube detector will consist of optical sensors deployed at depth into 
the thick Polar ice sheet. The ice will serve as Cherenkov medium for 
secondary particles produced in neutrino interactions in or around the 
instrumented volume.  
The successful deployment and operation of the 
AMANDA detector have shown that the Polar ice sheet is a suitable medium for
a large neutrino telescope and the analysis of the AMANDA data proves
the science potential of such a detector. 
 
IceCube will offer 
great advantages over AMANDA beyond its larger size: It will have a higher
efficiency and a better angular resolution in reconstructing tracks, it will
map electromagnetic and hadronic showers ({\it cascades}) from 
electron- and tau-neutrino interactions 
and, most important, it will have a superior energy resolution. 
Simulations, backed by AMANDA data, indicate that 
the direction of muons can be determined with sub-degree accuracy and
their energy measured to better than 30\% in the logarithm of the
energy. 
The direction of electron neutrinos that have produced electromagnetic 
cascades will be reconstructed to better
than 25$^\circ$  and the response in energy is linear with a 
resolution better  
than 10\% in the logarithm of the energy \cite{PDD}.
Energy resolution is critical because no atmospheric muon or neutrino 
background exceeds  1\,PeV in a cubic-kilometer detector
and full sky coverage is achieved. 

IceCube will  be able to 
investigate a large variety of scientific questions in 
astronomy, astrophysics, cosmology and particle physics \cite{PDD,Halzen2001}.
In this paper we will focus on the  
IceCube performance in searching for TeV to PeV muon neutrinos,
as expected from sources such as
active galactic nuclei, gamma-ray bursts or other cosmic accelerators 
observed as TeV gamma ray emitters.
We will present the results of a Monte Carlo study, including 
the simulation of the detector and 
the full analysis chain from filtering of the triggered data to
event reconstruction and selection.  
We will assess basic detector parameters like the pointing
resolution and the effective area of the detector directly from simulated data.
Further, we present a detailed calculation of the sensitivity of the detector 
to both diffuse and point-like neutrino emission following generic energy 
spectra, providing a benchmark sensitivity for some of the 
fundamental goals of high energy neutrino astronomy.

\section{The IceCube Detector}
\begin{figure}[htp]
\centering 
\epsfig{file=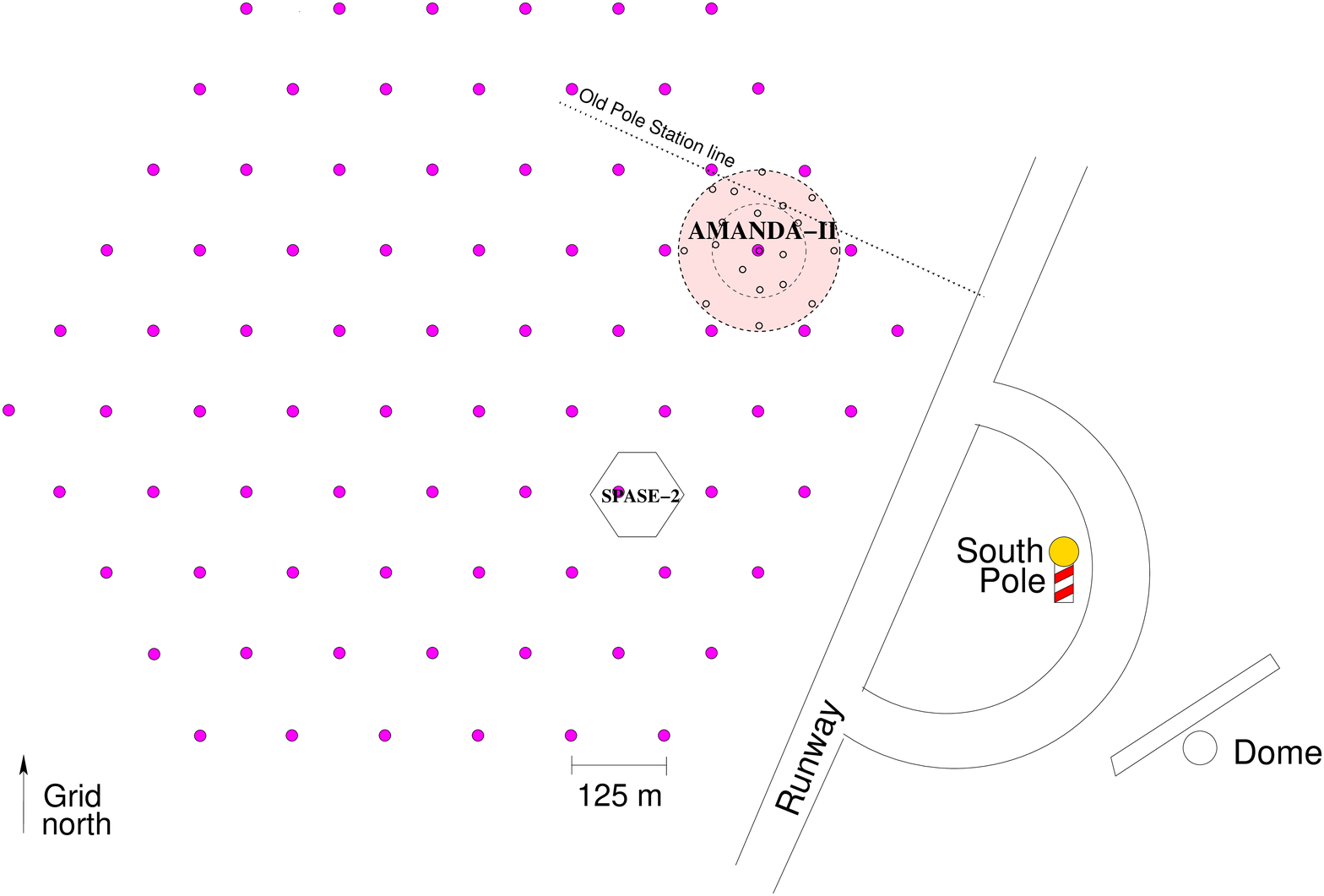,width=0.95\textwidth}
\caption[1]{
\label{fig:layout}Schematic diagram of the arrangement of the strings of the
\ic~ detector at the South Pole station. The existing AMANDA detectors and the SPASE 
air shower array will be embedded in the new detector. 
}
\end{figure}

The IceCube detector is planned as a cubic kilometer successor to
the AMANDA detector, currently operated at the South Pole.
It will consist of 4800 photomultiplier tubes (PMTs) of 10 inch
diameter,
each enclosed in a transparent pressure sphere.
These optical modules (OMs) will be deployed into vertical holes  
drilled to a depth of 2400\,m with pressurized hot water. 
As a significant improvement over the AMANDA technology,
 each OM will also house the 
electronics to digitize  the PMT pulses, retaining the 
full waveform information \cite{PDD,azriel}.
The digitized signals will be transmitted by twisted pair
cables to the surface data acquisition system. 
Adjacent OMs will be connected via a separate twisted pair
cable, which makes possible a local level-one hardware 
trigger in the ice. Local triggers will be combined by surface
processors to form a global trigger.
Triggered events will be filtered and reconstructed on-line, and 
the relevant information will
be transmitted via satellite to the northern hemisphere.

The dark noise rate of an OM will be 300-500 Hz.
This low rate is due to the sterile and 
low temperature environment of the deep Antarctic ice.
The absorption length of light from UV to blue
varies between 50\,m and 150\,m, depending on depth.
This allows photons to be collected even from very distant
tracks.
In fact, many of the photons will be delayed by scattering in the ice, 
but the some trajectory information may  be retrieved from the 
dispersed-signal PMT pulse shapes.   
The single-pulse timing accuracy will be
about 5\,ns, the sampling 
period 2-3\,ns and the double pulse
resolution close to 6\,ns. 
The dynamic range of one PMT
will be about 200 photo-electrons per 15\,ns, and 
several thousand photo-electrons integrated over the 
full waveform from a distant light source.

A central feature of the system is automatic time and 
amplitude calibration, critical for a large array at
remote location. Each OM will be equipped with a set of bright
370 nm LED pulsers. The LED signals, an essential part of the calibration system, 
can be seen by OMs as far away as 200\,m from their source LEDs.

The OMs will be deployed on 80 
vertical strings, each carrying 60 OMs spaced by 17\,m. 
The instrumented volume will span a depth ranging 
from 1400\,m to 2400\,m below the ice surface.
The strings will be arranged in a triangular pattern such that 
the distances between each string and its six nearest neighbors are  
125\,m. This configuration is the result of an extensive 
optimization procedure
\cite{leu98,mydiploma}, covering arrays with the number of OMs ranging from
2400 (half the design number) to 9600 (twice the design number),
with equally spaced strings, nested sub-arrays of larger density,
and a variety of ``exotic'' configurations.
A schematic sketch of the spatial string arrangement is shown in  
figure \ref{fig:layout}.
The complete
detector will be operational perhaps as soon as 
five years after the start of construction,
but during construction all deployed strings will produce high quality data 
for the collaboration to analyze.

The IceCube array deep in the ice will be complemented by IceTop, a
surface air shower array consisting of a set of 160 frozen
water tanks. The tanks will be arranged in pairs, separated by a few meters,
on top of each IceCube string.  
IceTop is the logical extension
of the SPASE surface array 
\cite{bai} which 
already is a unique asset for AMANDA.
The  data of the air-shower parameters measured at the surface combined with 
the 
signal of the high energy muon component under-ice provide
a new measure for the primary composition of cosmic rays.
Furthermore, IceTop will serve as a veto for 
air-shower-induced background and  it will provide cross checks 
for the detector geometry calibration, absolute pointing accuracy and angular
resolution. In addition, the energy deposited by tagged muon
bundles in air-shower cores will be  an external source for energy
calibration.

\section{Simulation and Analysis Chain}
The science potential of a kilometer scale neutrino telescope has been
assessed in previous papers by convoluting the expected neutrino induced 
muon flux from various astrophysical sources with an assumed square
kilometer effective 
detector area \cite{jodil,Halzen2001,HR}. 
In this work
we use a full simulation of the event triggering,
reconstruction and data selection to assess the detector 
capabilities.
The simulation of the detector and the analysis of the Monte Carlo data
rely on the software packages presently provided by the AMANDA collaboration  
\cite{siegmund,Hunderticrc}. This  means that the software concepts and 
analysis techniques used here have 
proven capable  and
have been verified by real data taken with the AMANDA detector. 
However, a full simulation of the IceCube hardware 
was not possible with the present software:
The simulated data correspond to the original 
AMANDA read-out, which does not yield the PMT waveforms.
Advanced analysis methods which take advantage of this additional 
information were not applied and hence we may yield a conservative picture of
the IceCube performance.

A brief outline of the different stages in the used processing chain 
is given in the following sections.

\subsection{Event Generation}
 The backgrounds for searches for extraterrestrial neutrinos come from 
the decay of mesons produced from cosmic ray (CR) 
interactions in the atmosphere.
The decay products include both muons and neutrinos. 
The muons will be responsible for the vast majority of triggers, since they are
very penetrating and are thus capable to reach the detector depth.
Air-shower induced
events can be identified by the fact that they involve exclusively
down-going tracks and a comparatively small deposit of Cherenkov light 
in the detector, as
the muons will
have lost most of their energy upon reaching the detector.
However, an up-going track might  be faked if two uncorrelated 
air-showers produce time-coincident muons within the detector.
In fact, about three percent of all triggered events will be caused by
muons from two independent air-showers.

The simulation 
packages {\tt Basiev} \cite{basiev} and {\tt Corsika} \cite{corsika}
were used for the CR induced muon background.
Roughly 2.4 million events containing 
muon tracks from one single air shower  {\bf (Atm $\mu^{\mathrm{single}}$)} 
were simulated with primary
energies up to 10$^{8}$\,GeV.
High energy events as well as events that contain tracks close to the horizon
were oversampled, in order to
achieve better statistics at high analysis levels.  
In addition, we have simulated 
one million events  
containing tracks from two independent 
air-showers {\bf (Atm $\mu^{\mathrm{double}})$}.

The muons induced by atmospheric neutrinos {\bf (Atm $\nu$)}  
form a background  
over $4\pi$\,sr and up to very high energies.   
However, the energy spectrum of atmospheric neutrinos falls steeply like 
$dN_{\nu}/dE_{\nu} \propto E_{\nu}^{-3.7}$, whereas one expects an energy 
spectrum as hard as $E^{-2}$ from shock acceleration mechanisms in
anticipated cosmic TeV-neutrino sources.
Therefore, cosmic neutrino energies should extend to higher values and 
cause  more light in the detector than will atmospheric neutrinos. 
The amount of light observed in an event 
is therefore useful as a criterion
to separate 
high energy muons induced by cosmic neutrinos from those induced by 
atmospheric 
neutrinos.
An uncertainty in the flux of atmospheric neutrinos arises from the 
poorly known contribution of prompt decays of charmed mesons produced in 
the atmosphere \cite{Costa,gaisserhonda}. 

Neutrino induced events have been simulated with the program {\tt nusim} 
\cite{MyPhD}. 
Neutrinos are sampled from a $1/E$ spectrum and are later re-weighted 
to produce different energy spectra as required. The code includes a 
simulation of the neutrino propagation through the 
Earth, taking into account  absorption in charged current interactions as well
 as neutral current regeneration. 
The neutrino cross sections are calculated using the  MRSG \cite{MRSG} parton
distributions. The column density of nucleons to be traversed is 
calculated according to 
the Preliminary Reference Earth Model \cite{PREM}.
Muons that are produced in the rock beneath the detector are propagated to the
rock/ice boundary using the Lipari-Stanev \cite{SL91} muon propagation code.
In total we have simulated $7.4 \cdot 10^{5}$ events 
induced by neutrinos with primary energies up to $10^{8}$\,GeV.
The flatness of the simulated $1/E$ neutrino spectrum leads to an oversampling 
of events at high energies for most of the energy spectra investigated.   

For ``conventional'' flux of atmospheric neutrinos ({\it i.e.} the
component related to decays of pions and kaons) we apply the 
 prediction calculated by Lipari \cite{Lipari93}. For the prompt charm
 contribution we compare  two different estimates: The ``Recombination Quark
Parton Model'' ({\tt rqpm}) developed by Bugaev \ea~ \cite{rqpm} and the 
model by Thunman \ea~ ({\tt TIG}) \cite{TIG}. 
The latter model predicts a substantially smaller rate
and may serve as a 
lower limit for the prompt charm contribution.

For the  flux of  extraterrestrial neutrinos {\bf (Cosmic $\nu$)} we apply a 
generic $E^{-2}$ energy spectrum, as 
expected from shock acceleration.  We use a source strength 
of  
$E^{2}_{\nu}\times dN_{\nu}/dE_{\nu}=10^{-7}$\diffunit~ as a benchmark 
diffuse flux
of extraterrestrial neutrinos.
This is  
an order of magnitude below present experimental limits
set on the flux of muon neutrinos \cite{diffuse} or electron 
neutrinos \cite{cascades}. 

\subsection{Muon Propagation}
The propagation of muons through the ice is modeled with either the 
code by Lohmann, Kopp and Voss \cite{Lohmann} (for muon energies smaller than 
$10^{5.5}$\,GeV) or the code by Lipari and Stanev 
\cite{SL91} (for muon energies greater than 
$10^{5.5}$\,GeV).
These codes calculate the stochastic radiative and nuclear interactive 
energy losses along the muon track within or close to the detector instrumented volume.

The complete tracking of all Cherenkov photons produced by the muon and
associated stochastic radiative energy losses for each event would require an impractical 
amount of computing power. 
Therefore, the photon amplitudes and timing distributions at all
points in space from  both  a muon and an electromagnetic
cascade are pre-calculated and tabulated into fast lookup tables using the
{\tt PTD} \cite{PTT} software package. This simulation takes into account  the
scattering and  absorption properties of the ice as well as the response of the
PMT. 

\subsection{Detector Simulation}
The response of the entire array of PMTs is modeled with the 
detector simulation {\tt amasim} \cite{Hunderticrc,amasim}. The actual number of photons at 
an OM is found by sampling from
a Poisson distribution with a  mean amplitude computed by summing over all
contributing muons and cascades. The arrival times of these photons are
sampled from  the pre-tabulated distributions.
Noise photons are added assuming a
PMT noise rate of 500 Hz.
For the trigger we use a local coincidence filter, requiring 
5 local coincidences within a time window of 7\,$\mu$s,
where a local coincidence
is defined as the registration of at least two signals in two or more
PMTs within a group of five neighboring OMs on a  string. 

The detector geometry used in this simulation differs from the finalized 
design in
 the total number of strings (we have simulated a 75 string detector 
instead of 80),
the total number of OMs (4575 instead of 4800),
the  string length (960\,m instead of 1000\,m) and
the location of the detector 
center (which was simulated at 2000\,m, while it will lie at 1800\,m depth 
in the updated design).
The spatial arrangement of the strings in a 125\,m spaced triangular grid
is in accordance to what is 
presented in the previous section. 

\subsection{Event Reconstruction}
The triggered events are first filtered based on three  
fast ``first guess'' algorithms which use the
arrival times of the photons or the topology of OMs having
registered a photon signal (or ``hit''):
{\bf (1)} The {\it line fit} ($LF$) is based on a simple analytic $\chi^{2}$
minimization \cite{linefit}. It fits  the free parameters 
(vertex position and  velocity) of a hypothetical straight line trajectory  
to the one-dimensional projection of an observed pattern of hits.
{\bf (2)} The {\it dipole approximation} \cite{recopaper} is based on the hit topology: 
The sum of  
all unit vectors pointing from one hit to the next in time
gives a ``dipole vector'' $\vec{M}$.
The direction of $\vec{M}$  is correlated to the 
direction of the incoming track(s), while its absolute value
is a measure of the goodness of the approximation.
{\bf (3)} The {\it direct walk algorithm} ($DW$) \cite{recopaper}
posits as track hypotheses the  straight line connections between 
every two hits that have occurred in separate OMs with a time 
difference consistent with 
muon flight time between these two OMs.
Those track hypotheses that pass a  consistency check  with 
respect to the  complete hit
pattern of the event are combined to obtain an estimate of the track 
parameters.
 
Following this first guess methods, the events are reconstructed
using a full {\it maximum likelihood reconstruction} ($LR$) \cite{recopaper,recoos}.
The probabilities in the likelihood function follow the arrival time
distribution of photons emitted along a track as a function of distance
and angle of the track with respect to the OM. These distributions
have been obtained from a photon propagation
simulation. 
The reconstruction used here 
relies on the information carried by the {\em first} photon 
that arrives at the PMT. This corresponds to the original AMANDA read-out 
that only yields the timing information of the PMT pulses. 
\footnote{In contrast, the IceCube electronics will retain the full pulse shape. 
Detailed hit information  can be extracted from the integrated charge and the
peak structure of the pulse.
Future reconstructions will therefore profit from 
the additional information formed by 
consecutively arriving photons which were multiple-scattered and delayed
on their way from the muon track to the PMT.}

\section{Basic Performance Capabilities}
\label{sec:performance}
The detector trigger rate for a five-fold local coincidence trigger was found
to be  1.7\,kHz.  This includes a 50\,Hz rate of triggers due to
non-correlated time-coincident air-showers {\bf (Atm $\mu^{\mathrm{double}}$)}.
As described below, a basic set of standard event selection criteria
was established that remove the bulk of the
down-going CR induced muons,
but still yield a large passing rate for muons from 
atmospheric neutrinos. These atmospheric neutrinos would then form the 
background to searches for cosmic neutrinos.
We use this level of data reduction as a baseline performance measure.

\subsection{Event Selection}
The most vital
criteria to reject the background of downgoing CR induced muons
are the zenith angles obtained from various reconstruction and 
filter algorithms
($\Theta_{LR}$, $\Theta_{LF}$, $\Theta_{DW}$ and $\Theta_{\vec{M}}$).
The easiest way to reject CR muons would be to select exclusively 
up-going tracks.
However, muons from PeV or EeV neutrino interactions are expected to arrive 
from directions close to or above the horizon, so it 
is worthwhile to combine the angular cut with an energy criterion.
If the neutrino interaction 
does not occur far from the detector, the energy deposit of the daughter muon 
will be high enough to distinguish it from
low energy CR muons. An estimator of this energy deposit is the number 
of OMs (or ``channels'') that have registered a hit.
We therefore accept downgoing tracks provided the channel multiplicity 
(\Nch) of 
the event is sufficiently large. 

The individual selection criteria are listed in table 
\ref{tab:cuts}.
The first three criteria  are based on the track directions obtained 
from  the three 
``first guess'' methods and aim at the early rejection of 
low energy downgoing CR muons. The level of data reduction 
achieved with the application of cuts 1--3  will be referred to as 
``level\,1''.

``Level\,2'', defined by cuts 4--9, is based on the 
observables of the more accurate (and more CPU intensive) 
likelihood reconstruction:

\begin{itemize}
\item{
Events that are reconstructed 
with zenith angles smaller than $85^\circ$ 
({\it i.e.} directions 
more than five degrees above the horizon) are rejected, 
as long as \Nch~ is smaller than 150.
The \Nch-criterion is tightened with decreasing zenith angle
($\Theta_{LR}$) [cut 4].
}
\end{itemize}
Apart from the direction criterion, the likelihood reconstruction provides a
series of quality parameters, 
which we apply  in order to select a  
sample of high quality and well reconstructed events:
\begin{itemize}
\item{
We require the
{\it reduced likelihood} {\bf ($\mathcal{L}$)} of the likelihood reconstruction
to be sufficiently small.
$\mathcal{L}$ is given by the negative
logarithm of the likelihood of the best-fit track hypothesis
divided by the number of degrees of freedom of the fit, hence a {\em small} value indicates a good track 
quality [cut 5].
}
\item{We require a minimum 
{\it number of direct hits} {\bf ($N_{\mathrm{direct}}$)}, {\it i.e.} hits  that have
occurred with a sufficiently small delay ($<150$\,ns) 
relative to the arrival time predicted for an unscattered Cherenkov
photon emitted from the reconstructed track
[cut 6].
}
\item{ 
We require a minimum {\it track length ($L$)}, {\it i.e.} a minimum distance along
the reconstructed track over which the hits were detected. We define this length as 
the maximum distance of two hit positions projected on the straight line defining the track direction.
A more stringent criterion is a lower bound on the track length  
using direct hits {\bf ($L_{\mathrm{direct}}$)} 
[cut 7].
}
\item{ 
The consistency of the fitted track direction is checked with the
{\it  smoothness  parameter} \cite{atmospheric,recopaper}. 
It is a measure of the
evenness of the projection of the hit positions along the track, based on a
 Kolmogorov-Smirnov test.
The smoothness parameter can
be calculated with either all hits {\bf($S$)}, or exclusively direct hits 
{\bf ($S_{\mathrm{direct}}$)} 
[cut 8].
}
\item{
For high quality tracks, the various reconstruction methods are likely to 
produce similar 
results close to the original ({\it i.e.} true) track. 
We therefore require the difference in
zenith angles obtained by two different methods to be small
[cuts 9 and 3].
}
\end{itemize}

\begin{table}[htb]
{\footnotesize
 \begin{center}
   \caption{{\bf Definition of cuts levels}.}
   \vspace{0.2cm}
    \begin{tabular}{rccl}
      \hline
       & {\bf Parameter}  & {\bf Cut} & \multicolumn{1}{c}{{\bf Explanation}}\\
       \hline
       \hline 
       \multicolumn{4}{l}{\it Level 1:}\\
       \hline
       \hline 
       1.  &  $\Theta_{LF}$ &  $>60 ^\circ$ if $N_{\mathrm{ch}}<50$ & 
       zenith angle criterion \\[-.35cm] 
       & & & based on  $LF$, applied for \\[-.35cm] 
       & & & low multiplicity events \\
      \hline     
       2.  & $\Theta_{\vec{M}}$ & $>50^\circ$  if $|\vec{M}|>0.2$ & 
       zenith angle criterion \\[-.35cm] 
       & & & based on $\vec{M}$, applied for  \\[-.35cm]
       & & &  high goodness-of-fit values \\
      \hline 
       3.  & $|\Theta_{DW} - \Theta_{\vec{M}}|$  & $< 50^\circ$ & 
       consistency of $LF$ and $DW$ \\
      \hline
      \hline
      \multicolumn{4}{l}{\it Level 2:}\\
      \hline
      \hline 
       4.  & $\Theta_{LR}$  &  $> 85^\circ$  & zenith angle criterion of $LR$\\*[-.35cm] 
       & & {\it or}  & which is weakened with \\[-.35cm]
       & &  $N_{\mathrm{ch}} > 150+250*\cos(\Theta_{LR})$ & 
       increasing channel multiplicity\\
      \hline
       5.  & $\mathcal{L}$ &  $< 10$ & reduced likelihood of $LR$\\
      \hline
       6.  & $N_{\mathrm{direct}}$ & $ > 10$  if  $N_{\mathrm{ch}}<50 $ & 
       requirement of 10 direct hits \\[-.35cm] 
        & & & for low multiplicity events \\
      \hline
       7.  & $L$  & $ > 300\,{\mathrm m}$ & requirement of minimum 
       track\\*[-.35cm]
           & & {\it and} & 
       length, using direct hits for\\[-.35cm]
       &  $L_{\mathrm{direct}}$ & $> 300$\,m  if $N_{\mathrm{ch}}<150$ &  multiplicities smaller than 150\\
      \hline
       8.  & $|S|$ &  $< 0.5$ & constancy of light output along\\[-.35cm]
          &  & {\it and}  & 
      the track, requirement is\\[-.35cm]
      & $|S_{\mathrm{direct}}|$  &  $< 0.5$  if $N_{\mathrm{ch}}<50 $ & tightened for low multiplicities\\
      \hline
       9.  & $|\Theta_{LF} - \Theta_{LR}|$ &  $< 10^\circ$  if  $N_{\mathrm{ch}}<150$ & 
       consistency of $LF$ and $LR$\\
      \hline
    \end{tabular}
    \\*[1.cm]
    \label{tab:cuts}
  \end{center}
}
\end{table}

~\\
These quality criteria are particularly important for
muons that travel a  short path through the instrumented detection volume, {\it e.g.}
low energy muons or muons that pass 
only through  
the outer rim  of the detector or even outside its geometrical volume.   
These muons will cause signals in fewer OMs and therefore
leave less information for the reconstruction. Most of the quality criteria are
therefore tightened if the channel multiplicity (\Nch) 
observed in the event is 
small. 

~\\

\subsection{Muon Detection Rates}

\begin{table}[htb]
 \caption{{\bf Passing rates} for signal and background Monte Carlo events. 
The signal expectation corresponds to a source flux of $ E^{2}_{\nu} 
\times dN_{\nu}/dE_{\nu}$ = 10$^{-7}$\,\diffunit. 
The expectation for atmospheric neutrino events 
is listed separately for the ``conventional'' component  and the
``prompt'' component (following \protect\cite{TIG}({\tt TIG}) and 
\protect\cite{rqpm}({\tt rqpm})). 
The fraction of
prompt charm events  with respect to the
 whole atmospheric neutrino sample is given in parentheses.
The numbers of cosmic ray muon background events 
are shown separately for events that contain  
muon(s) from only one air shower (Atm $\mu^{~\mathrm{single}}$) and those that
 contain muons 
from two accidentally coinciding air showers (Atm $\mu^{~\mathrm{double}}$).  
The errors are statistical only.}

\label{tab:passclasses}
{\scriptsize
\renewcommand{\arraystretch}{1.2}
  \begin{center}
~\\[0.2cm]
   \begin{tabular}{crlrlrl}
    \hline
    ~           &  Trigger        &~                   & Level 1 &~                 &   Level 2 &~    \\
    \hline \hline
    {\bf Cosmic $\nu$} & 3\,331 $\pm$ 6 &~& 2\,172 $\pm$ 4 &~  &  1\,089 $\pm$ 3 &~\\
    \hline
    {\bf Atm $\nu$}       &  (824 $\pm$ 4) &   $\times 10^3$   & (264 $\pm$ 2) & $\times 10^3$ & (91 $\pm$ 1) & $\times 10^3$ \\[0.2cm]
      TIG                 &  (0.97  $\pm$ 0.003) & $\times 10^3$ & (0.40   $\pm$ 0.002) & $\times 10^3$  & (0.17 $\pm$ 0.001) & $\times 10^3$ \\
      ~ & \multicolumn{2}{c}{(0.1\,\%)}  &   \multicolumn{2}{c}{(0.2\,\%)}   &  \multicolumn{2}{c}{(0.2\,\%)} \\[0.2cm]
 
      rqpm           &  (24.8  $\pm$ 0.07)  & $\times 10^3$      & (11.08 $\pm$ 0.04)  & $\times 10^3$   &  (4.85 $\pm$ 0.03)  & $\times 10^3$  \\
      ~              &   \multicolumn{2}{c}{(3\,\%)}   &   \multicolumn{2}{c}{(4\,\%)}     &    \multicolumn{2}{c}{(5\,\%)} \\
    \hline
    {\bf Atm $\mu^{~\mathrm{single}}$} & (5.2  $\pm$ 0.01) & $\times 10^{10}$ & (1.3 $\pm$ 0.01) & $\times 10^{9}$  & (72 $\pm$ 3)  & $\times 10^3$ \\ 
    {\bf Atm $\mu^{~\mathrm{double}}$} & (1.6 $\pm$ 0.02) & $\times 10^{9}$   & (4.6  $\pm$ 0.3) & $\times 10^{7}$ & (28 $\pm$ 7)  & $\times 10^3$ \\
    \hline
    \end{tabular}
   \\*[1.cm]
   \end{center}
}
\end{table}

We compare the detector response as well as the
event selection efficiency for all types of events:
CR muons, muons induced by atmospheric neutrinos and
muons from cosmic neutrinos with a hard energy spectrum,  following an 
$E^{-2}$ power law.
The numbers of triggered and selected
events at each level, normalized to one year of 
data taking, are listed in table 
\ref{tab:passclasses}.  
With a flux of
$E_{\nu}^{2} \times dN_{\nu}/dE_{\nu}$ = 10$^{-7}$\,\diffunit~
adopted as a benchmark for the flux
of cosmic neutrinos, we 
expect more than 1000 signal events at level\,2.
At this stage, both the background from atmospheric neutrinos and the 
background from CR muons yield roughly $10^{5}$ events per year.
The {\tt rqpm} model for atmospheric charm predicts a contribution 
of almost 5000 prompt charm events to the atmospheric 
background. The {\tt TIG} model predicts thirty times fewer events.
\begin{figure}[htp]
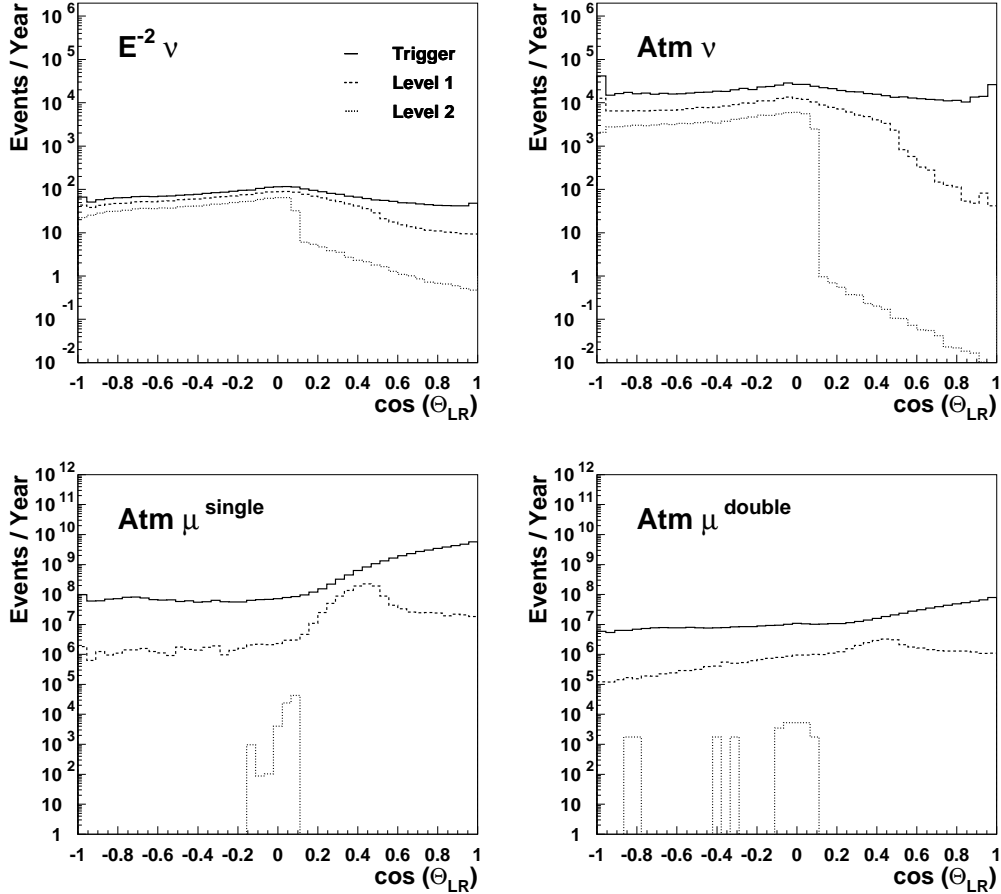

\begin{center}
\epsfig{file=./plots/zen1.pass.FETTBW.epsi,
width=.45\textwidth
}
\hskip .5cm
\epsfig{file=./plots/zen0.pass.FETTBW.epsi,
width=.45\textwidth
}
\vskip .5 cm
\epsfig{file=./plots/zen8.pass.FETTBW.epsi,
width=.45\textwidth
}
\hskip .5cm
\epsfig{file=./plots/zen9.pass.FETTBW.epsi,
width=.45\textwidth
}
\end{center}
\caption{{\bf Distribution of the reconstructed zenith angle for different 
event classes:}
Signal from a diffuse flux of cosmic neutrinos following an $E^{-2}$ 
spectrum ({\it top left}), atmospheric neutrino 
background including charm according to \protect\cite{rqpm} ({\it top right}),
atmospheric muon background from one single air-shower ({\it bottom left}) 
and two coincident air-showers
({\it bottom right}). The individual histograms in each plot correspond to: 
Trigger
level (full lines), samples after applying level\,1 and level\,2 cuts 
(dashed and dotted lines, respectively). Event numbers are normalized to one
year.
\label{fig:zenrej}
}
\end{figure}

Figure \ref{fig:zenrej} shows the zenith angle distribution 
for the reconstructed zenith angle $\Theta_{LR}$ of the 
four different event classes ({\bf Cosmic $\nu$}, 
{\bf Atm $\nu$},
{\bf Atm $\mu^{\mathrm{single}}$} and {\bf Atm $\mu^{\mathrm{double}}$})
at different cut levels.
The level\,1 selection removes the bulk of low energy 
down-going CR-induced background. The angular cut on the zenith angles
of the ``first-guess'' methods is still soft, so that most of the remaining 
background is  located in the region around 30$^\circ$ above the horizon.
Level\,2 then restricts the allowed zenith region to 5$^\circ$ above the 
horizon, except for very bright, high multiplicity events.   
The remaining ordinary CR muon background ({\bf Atm $\mu^{\mathrm{single}}$}) 
at level\,2 is then concentrated 
at the horizon and could be rejected with a tightened cut
on the zenith angle, while the sample of 
CR-induced background composed of two air-showers ({\bf Atm $\mu^{\mathrm{double}}$})
still contains
misreconstructed events that ``fake'' an upward track. 
However, filter level\,2 does not contain a definite energy 
selection yet, required to 
separate the  high energy signal of cosmic neutrinos 
from the atmospheric neutrino background. 
In the simplest approach this energy selection is an additional  tight cut
on the channel multiplicity.
This final cut has to be optimized for different 
analysis purposes (see section \ref{sec:diffsens}), 
but in any case it will lead to a drastic reduction of all 
three classes of background.
In this analysis none of the CR muon events did pass this additional cut.

\begin{figure}[htp]
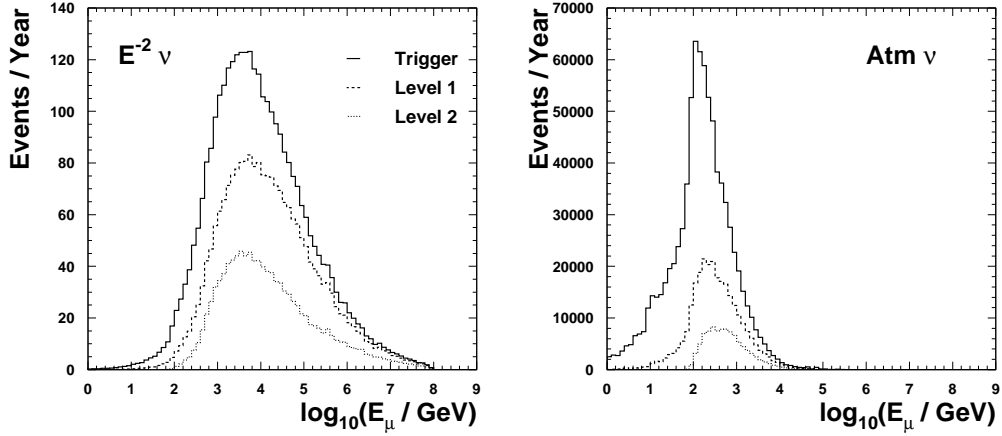

\begin{center}
\epsfig{file=./plots/muenspectr.1.liny.epsi,
width=.45\textwidth
}
\hskip .5cm
\epsfig{file=./plots/muenspectr.0.liny.epsi,
width=.45\textwidth
}
\end{center}
\caption{{\bf Energy spectra of neutrino induced muons at different cut-levels:}
Signal from a $E^{-2}$-source ({\it left}), atmospheric
neutrino background ({\it right}). The individual histograms in each plot correspond to: Trigger
level (full lines), samples after applying level\,1 cuts (dashed lines) and
level\,2 cuts (dotted lines). The energy spectra show the energy at the 
point of closest approach to the detector center. 
\label{fig:enrej}
}
\vskip 1.cm
\end{figure}

\begin{figure}[htp]
\begin{center}
\epsfig{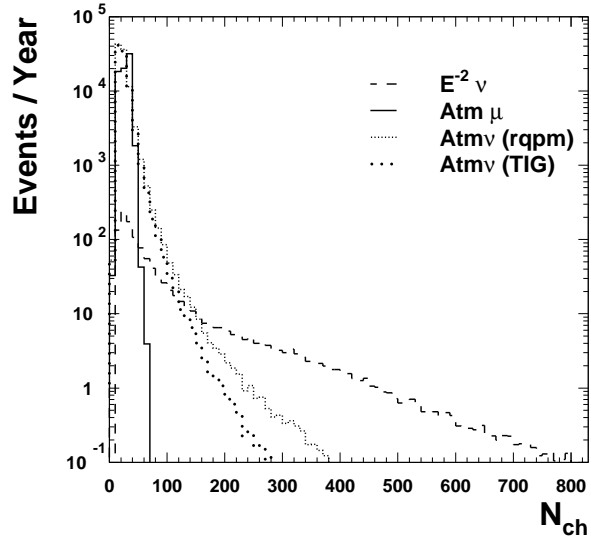}
\end{center}
\caption{{\bf Channel multiplicity for different event classes at level\,2:} 
For signal from $E^{-2}$ source (dashed), atmospheric neutrinos including 
the two alternative charm contributions {\tt TIG} (sparse dots) and {\tt rqpm} 
(dense dots) and CR muon events (full lines).
\label{fig:enrej2}
}
\vskip 1.cm
\end{figure}

The energy spectra of muons generated by cosmic and atmospheric neutrinos
are shown in  figure \ref{fig:enrej}. 
At the point of their closest approach to the detector center, muons from a
cosmic  $E^{-2}$ neutrino source typically have energies in the TeV-PeV region,
whereas the background of muons induced by atmospheric neutrinos peaks between
 100 and 300\,GeV. Figure \ref{fig:enrej2} shows the channel multiplicity 
distributions of all event classes at level\,2. The signal of high energy cosmic neutrinos 
shows a clear excess at high multiplicities compared to the lower energy background events. 

\subsection{Effective Detector Area}

\begin{figure}[htp]
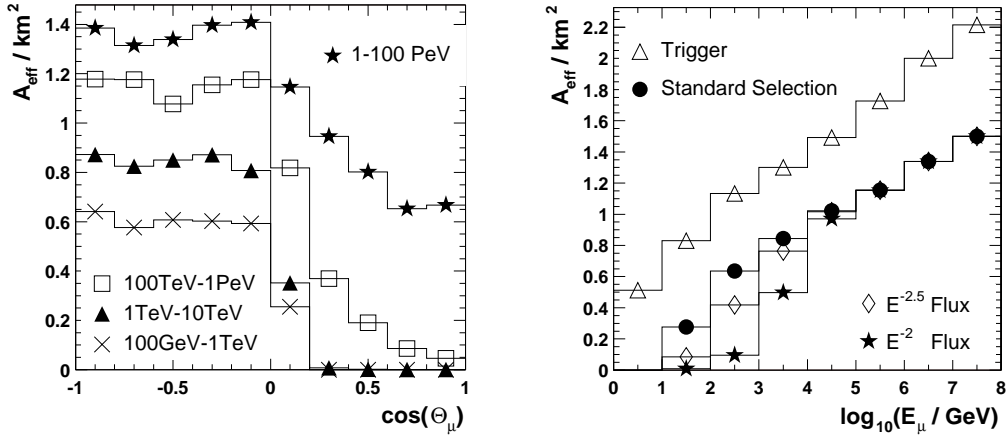

\begin{center}
\mbox{}
\vskip .4cm
\epsfig{file=./plots/aeff_zenith.dependent.cuts.epsi,
width=.44\textwidth
}
\hskip .9cm
\epsfig{file=./plots/aeff_energy.depend.epsi,
width=.44\textwidth
}
\end{center}
\caption{{\bf Effective area} as a function of the zenith angle after applying level\,2 cuts
({\it left}) and as a function of the muon energy for different cut levels 
as described in the text
({\it right}).
The zenith dependence is shown separately for different 
intervals of $E_{\mu}$.
The energy dependence holds for muons for which $\cos \Theta_{\mu} < 0$, 
{\em i.e.} muons that arrive from the northern sky.
Energy cuts adjusted to signal spectra following  $E^{-2}$ and $E^{-2.5}$
power laws change
the energy threshold. 
\label{fig:aeff.en}
}
\vskip 1.cm
\end{figure}

As a measure of the detector efficiency we use the effective detector area, defined as 
\begin{equation}
    A_{\mathrm{eff}}(E_{\mu},\Theta_{\mu} ) =
  \frac{N_{\mathrm{detected}}(E_{\mu}, \Theta_{\mu})}{N_{\mathrm{generated}}(E_{\mu},\Theta_{\mu} )} \times A_{\mathrm{gen}},
\end{equation}
where
$N_{\mathrm{generated}}$ is the number of muons in the test sample that have 
an energy
$E_\mu$ at a given point within the fiducial volume  
and an incident zenith angle $\Theta_{\mu}$. 
In the following we give $E_\mu$ at  the point of closest approach 
to the detector center (which might lie outside the geometrical
detector volume).
$N_{\mathrm{detected}}$ is 
the number of events that  trigger the detector  or pass the cut level
under consideration. The fraction of generated to triggered or selected
events is scaled with the size of the generation plane $A_{\mathrm{gen}}$,
which is the cross-sectional
area of the cylinder that contains all generated muon tracks with directions 
parallel to its axis.

The left plot in figure \ref{fig:aeff.en} shows the effective area
at cut level\,2 as a function of the zenith angle of the muon tracks.
Results are shown for four separate energy intervals.
The detector will have an effective detection area of one square kilometer
for upward moving muons in the  TeV range. Above 100\,TeV   
the selection allows for a detection of downgoing neutrinos,  
{\it i.e.} for an observation of the southern  
hemisphere ($\cos \theta_{\mu} > 0$). In the PeV range the effective area
for downgoing muons is above 0.6 km$^{2}$, increasing towards the horizon.
This means that IceCube can  observe a large part of our Galaxy, including the
Galactic center.

The right plot in figure \ref{fig:aeff.en} shows the effective area
as a function of the muon energy at closest approach to the detector center. 
Here, the effective area was calculated using a sample of muons which arrive 
from the lower hemisphere, {\em i.e.} using tracks with incident zenith angles 
larger than $90^{o}$ (or $\cos \theta_{\mu} < 0$, accordingly).
In that sense the numbers indicate average values for the 
northern hemisphere.  
At trigger level the detector shows a sizeable acceptance even for low energy events.

The effective trigger area  
reaches one square kilometer at a few hundred GeV. 
Roughly 50\% of all triggered events 
pass the ``standard selection'' (level\,2),
independent of the muon energy.
Best sensitivities  to extraterrestrial neutrinos are obtained 
by increasing the energy threshold by additional cuts on 
the muon energy. 
The optimal threshold, {\it i.e.} the threshold where optimal sensitivity 
to one particular signal is 
achieved, is determined by the signal shape. 
For instance, a hard signal 
spectrum like $E^{-2}$ would 
suggest a tighter cut than a  softer spectrum 
falling like $E^{-2.5}$.  
In order to show this effect the effective area was calculated 
after applying different energy cuts. In addition to 
trigger level and  level\,2, 
figure \ref{fig:aeff.en} shows the effective area after applying
additional  
energy cuts optimized for probing hypothetical point sources
with signal spectra following  $E^{-2}$ and $E^{-2.5}$
(see section \ref{sec:point.sens}).

\subsection{Angular resolution}
The angular resolution of the detector is an important quantity for the search 
for  neutrinos from point sources. 
A good angular resolution allows for a small search bin,
 resulting in a low background rate per bin. 
We characterize the angular resolution by 
the median of the 
distribution of the space angles between 
the true and reconstructed directions of the simulated muon tracks.

\begin{figure}[htp]
\begin{center}
\epsfig{file=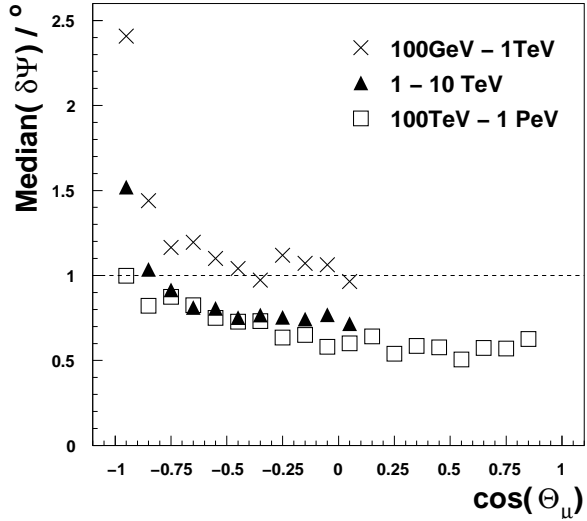,
width=.55\textwidth
}
\end{center}
\caption{{\bf Pointing resolution for neutrino-induced muon events.}
Shown is the median space angle error of the likelihood reconstruction 
as a function 
of  the zenith angle of the incident track. The median was calculated for
 an energy spectrum $\propto\,E^{-2}$ and after applying level\,2 cuts.
\label{fig:pointing}
}
\vskip 1.cm
\end{figure}

The angular resolution after applying level\,2 cuts is shown
 in figure \ref{fig:pointing} 
as a function of the zenith angle of the  muon tracks.
In the energy region from 100\,GeV to 1\,TeV the pointing 
resolution approaches 
$1^\circ$ for tracks with zenith angles smaller than $140^\circ$. 
For nearly vertical tracks of low energy muons 
the angular resolution is worse,
because these
events are likely to cause hits in optical modules of only a single string. 
However, the reconstruction accuracy in this energy range is 
similar to the mean angle between the muon and the initial neutrino.  
In the most promising energy range, the region   
at few TeV and above, the resolution is 
substantially improved. Also, the zenith angle dependency of the pointing 
resolution weakens towards
higher energies.
Most of the signal in the TeV -- PeV region will be reconstructed 
with an accuracy significantly better than  $1^\circ$.
Current reconstruction methods achieve a resolution 
close to $0.5^\circ$ for events near the horizon.
However, we expect significant improvement of the pointing resolution with further 
development of the reconstruction, 
in particular from including amplitude and waveform information.

\clearpage
\newpage

\section{Sensitivity to Astrophysical Sources of Muon Neutrinos}
In most theoretical models, the production of high energy 
CR is accompanied by the production of mesons.
Prominent candidates for CR sources are putative cosmic accelerators
like active galactic nuclei (AGN), microquasars, supernova remnants
and gamma ray bursts (GRBs). Theoretical models for 
such objects usually involve shock acceleration of protons. The
protons  interact with ambient matter or radiation fields
producing mesons that subsequently decay into neutrinos.    
The spectral distribution of neutrinos expected from cosmic accelerators is 
 $dN_{\nu}/dE_{\nu} \propto E_{\nu}^{-2}$, or even harder,
depending on the predominant meson production mechanism in the source
and on full particulars of the acceleration.

The sum of all cosmic accelerators in the universe should 
produce an isotropic flux of high energy neutrinos, which
would be observable as an excess above the diffuse flux of atmospheric 
neutrinos.
The absolute flux of  individual sources may be small, and requires 
a careful selection, if one wants to resolve it.
However, in this case, one can strongly suppress background,
because the number of background events will be reduced
with the size of the spatial 
search bin or -- in case of transient phenomena -- the duration
of the observation time window.

In the following we calculate the sensitivity for diffuse fluxes of cosmic
muon neutrinos as well as for fluxes from individual point sources, both 
steady and transient (GRBs). In contrast to former analyses, 
which were
based on simple assumptions on the detector effective area as well as on 
its energy 
resolution \cite{jodil,Halzen2001,HR}, 
the method we apply involves exclusively event observables  that
will be available from real data taken by \ic. 

\subsection{Calculation of the Sensitivity}
We explore the sensitivity of the IceCube detector to cosmic neutrino  
fluxes in two ways. First we consider the limits that would be placed on
models of neutrino production if no events were to be seen above those
expected
from atmospheric neutrinos. Second, we evaluate the 
level of source flux required to observe an excess 
at a given significance level.

\subsubsection{Limit setting potential}
Feldman and Cousins have proposed
a method to quantify the ``sensitivity''  of an experiment
independently of experimental data by calculating the average upper limit 
$\bar{\mu}$ that would be obtained in absence of
a signal \cite{fc98}.  
It is calculated from the mean number of expected background 
events,  $\langle n_{b} \rangle$, by averaging over all limits obtained   
from all possible experimental outcomes.
The average upper limit
is the maximum number of events that
can be excluded at a given confidence level. That is, the experiment can be expected
to constrain any hypothetical signal that predicts at least
$\langle n_s \rangle = \bar{{\mu}}$ signal events.    

From the 90\% c.l. average upper limit, we define the ``model rejection factor'' ({\it mrf})   
for an arbitrary source spectrum $ \Phi_{s}$ predicting
$\langle n_s \rangle$ 
signal events, as the ratio of the average upper limit 
to the expected signal \cite{HR}.
The average flux limit $\Phi_{90}$  is found by scaling 
the normalization of the flux model $\Phi_s$
such that the number of expected events 
equals the average upper limit
 
\begin{equation}
\label{scaleflux}
   \Phi_{90} ~= ~\Phi_s \times \frac{\bar{\mu}_{90}}{\langle n_s \rangle} 
   ~\equiv~\Phi_s \times  ~\mbox{\it{mrf}}.
\end{equation}

\subsubsection{Discovery potential}

For our purposes, a phenomenon is considered ``discovered'' when a measurement 
yields 
an excess of 5-sigma over background, meaning that the probability of the 
observation being due to an upward fluctuation of background is less than
$2.85 \times 10^{-7}$, being the the integral of the one-sided tail beyond 5-sigma 
of a normalized Gaussian. 
From the background expectation $\langle n_b \rangle$,
we can determine the minimum number of events  $n_0$ to be observed 
to produce the required significance as

\begin{equation}
\sum_{n_{\mathrm{obs}} = n_{0}}^{\infty} P(n_{\mathrm{obs}} ~|~ \langle n_{b} \rangle) 
\leq 2.85 \cdot 10^{-7}, 
\end{equation}
where $P(n_{\mathrm{obs}} | \langle n_b \rangle)$ is the Poisson probability 
for observing  
$n_{\mathrm{obs}}$ background events.
The minimum detectable flux $\Phi_{5\sigma}$ for any source model can then be 
found by scaling the model flux $\Phi_s$ such that 
$\langle n_s \rangle + \langle n_b \rangle = n_0$.  

If a real signal source of average strength $\Phi_{5\sigma}$ is present, the
probability of the combination of signal and background producing 
an observation 
sufficient to give the required significance
({\it i.e.}  an observation of $n_0$ events or greater) is

\begin{equation}
P_{5 \sigma} ~=~ \sum_{n_{\mathrm{obs}} = n_{0}}^{\infty} P(n_{\mathrm{obs}} ~|~ \langle n_{s} \rangle + \langle n_{b} \rangle).
\end{equation}

Thus we cannot say that an underlying signal strength will always produce 
an observation with 5-sigma significance, but we can find the signal 
strength such that the probability of $P_{5\sigma}$ is close to certainty
{\it e.g.} 70\%, 90\% or 99\%.

\subsection{Diffuse Flux Sensitivity}
\label{sec:diffsens}
Many models have been developed that predict a diffuse neutrino flux to 
be expected 
from the sum of all active galaxies in the universe. First 
we
will consider the potential of IceCube to both  place a limit on, and
detect, a generic diffuse flux following an $E^{-2}$ spectrum.
After looking in detail at this case we summarize the capabilities of the
detector to place limits on a few participating models with spectral shapes
different from $E^{-2}$.

\begin{figure}[htp]
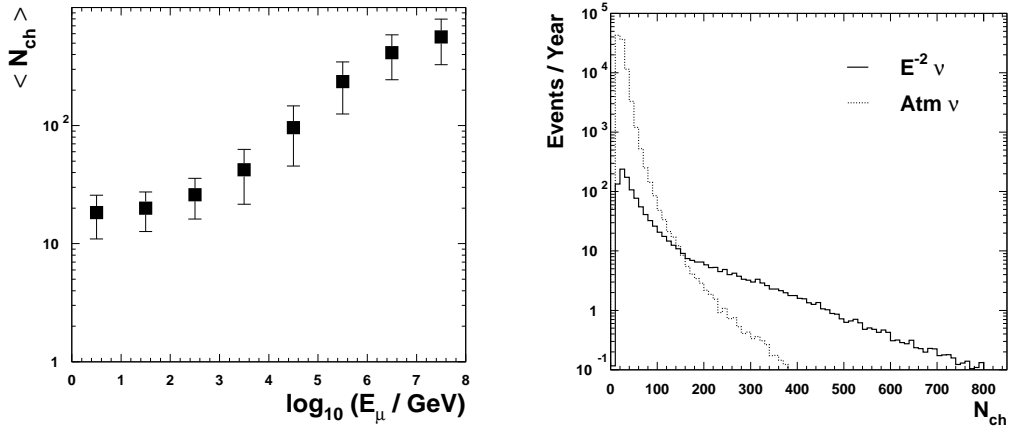

\begin{center}
\epsfig{file=./plots/mean.nch.vs.en.21.epsi,width=.44\textwidth}
\hskip .9cm
\epsfig{file=./plots/diff.orig.def.1.epsi,width=.44\textwidth}
\end{center}
\caption[1]{
\label{nchvsneutrinoenergy} {\bf Channel multiplicity.}
{\it Left:} Correlation between muon  
energy at closest approach to the detector center and detected channel 
multiplicity. 
Data points show to the mean number of OMs with a signal, averaged over one decade in energy. 
Error bars indicate the spread of the corresponding distribution.
{\it Right:} Distribution of detected channel multiplicity for 
$E^{-2}$ signal and atmospheric background.  
}
\vskip 1.cm
\end{figure}

We use the simplest observable, the multiplicity $N_{\mathrm{ch}}$
 of hit channels per event  
as an energy separation
cut, in order to reject the steep spectrum of events induced by atmospheric 
neutrinos,
and
retain the events from the harder extra-terrestrial diffuse 
spectrum.\footnote{An improved energy separation is expected from the use of 
a more sophisticated energy reconstruction using individual hit 
amplitude and/or the full waveform information}  
The correlation
between the muon 
energy at closest approach to the detector center
and channel multiplicity is shown in 
the left plot of figure \ref{nchvsneutrinoenergy}. The right plot shows
the \Nch~ distributions for an $E^{-2}$ signal compared to the atmospheric 
background.  

~\\

\begin{figure}[htp]
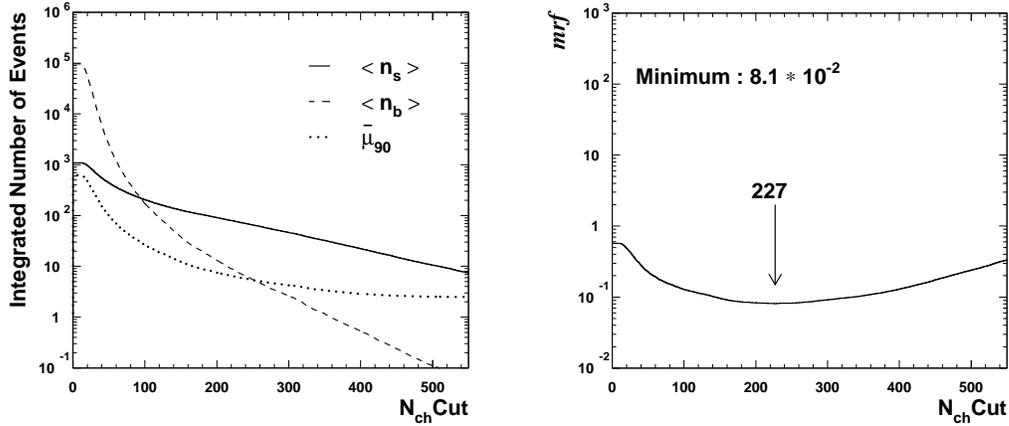

\begin{center}
\epsfig{file=./plots/diff.int.def.epsi,width=.44\textwidth
}
\hskip .9cm
\epsfig{file=./plots/mrf.diff.epsi,width=.44\textwidth
}
\end{center}
\caption{{\bf Cut optimization}.{\it Left:} Integrated distribution of 
detected channel multiplicity per event. {\it Right:} Model rejection factor 
with respect to a model source flux of
$\dNdE = 10^{-7}$\,\diffunit~ as a function of the applied \Nch-cut. 
\label{diffuse_integrated}
}
\vskip 1.cm
\end{figure}

In order to determine the \Nch-cut that achieves the best sensitivity
we optimize the cut with respect to  the 
model rejection factor ({\it mrf}) \cite{HR}. 
For each possible cut value we compute the {\it mrf}  from the 
number of remaining signal and background events.
The cut is placed, where the {\it mrf} is minimized.

\begin{figure}[htp]
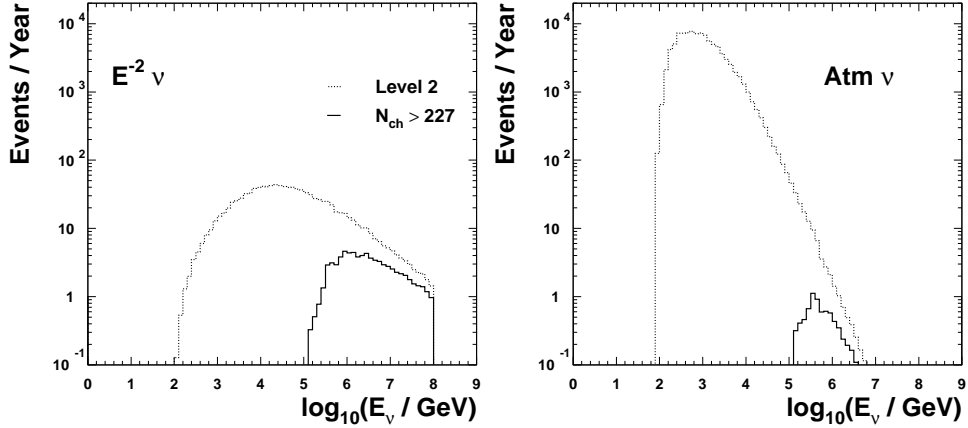

\begin{center}
\epsfig{file=./plots/nupass.trueen.diff.1.fin.selec.FETTBW.epsi,
width=.45\textwidth
}
\hskip .05cm
\epsfig{file=./plots/nupass.trueen.diff.0.fin.selec.FETTBW.epsi,
width=.45\textwidth
}
\end{center}
\caption{{\bf Energy spectra of selected neutrinos} for a $E^{-2}$
source ({\it left})
 and atmospheric neutrinos
({\it right}). The  selection is given by level\,2 cuts
(dotted  lines) and application of the optimized cut \Nch$>227$ (full lines).
The cutoff in the signal spectrum at $10^{8}$\,GeV is due to the limited 
energy range of simulation. 
\label{fig:nu.energy.spectra}
}
\vskip 1.cm
\end{figure}

This procedure is illustrated in figure \ref{diffuse_integrated}: The left 
plot shows the integrated multiplicity distribution for signal and background,
together with the average upper limit  $\bar{\mu}_{90}$. 
The right plot shows the {\it mrf} as a function of the \Nch-cut.
The 
{\it mrf} reaches its  minimum of $8.1 \times 10^{-2}$ at \Nch=227, 
corresponding to
an overall flux limit of  $E^2_{\nu} \times dN_{\nu}/dE_{\nu} = 8.1 \times 10^{-9}$\,\diffunit. 
This flux limit applies for the flux of extraterrestrial muon-neutrinos to be 
measured at the Earth. In the presence of neutrino oscillations, the constraint on the 
flux escaping cosmic sources must be modified accordingly: For maximal mixing \cite{superk,sno}
between muon- and tau-neutrinos during propagation to the Earth, one would expect the 
flux of muon neutrinos at the Earth to be half as large as the flux at the source. So the limit on the
muon neutrino flux {\em produced} in cosmic sources is higher by a factor of two.   
In the following ``cosmic neutrino flux'' refers to the intensity of muon neutrinos 
measured at the Earth.

The simulated cosmic neutrino flux of 
$E_{\nu}^{2} \times dN_{\nu}/dE_{\nu} = 1 \times 10^{-7}$\,\diffunit~\\  
yields 74 signal events passing the optimized cut,
compared to 8 background events from atmospheric neutrinos. 
The background expectation was calculated 
using the  {\tt rqpm} model for the prompt charm contribution, according to which  
prompt charm decays
account for 80\% of the remaining atmospheric neutrinos.
The prediction 
according to  the {\tt TIG} model
would result in an improvement of the 
average flux limit by roughly a factor of two.

The energy spectra of the incident signal and background neutrinos  
are shown in figure \ref{fig:nu.energy.spectra}. The final cut, as it is 
placed here, results in a detection threshold of about 100\,TeV.
This threshold is the result of an optimization to
one particular signal hypothesis, an $E^{-2}$ neutrino spectrum extending 
up to  energies of  $10^{8}$\,GeV. 
The artificial cutoff in the signal spectrum at 
$10^{8}$\,GeV (where the simulation ends) 
neglects additional events above this energy. Without the artifical cutoff one would get a slightly improved limit.

\begin{figure}[htp]
\centering
\epsfig{file=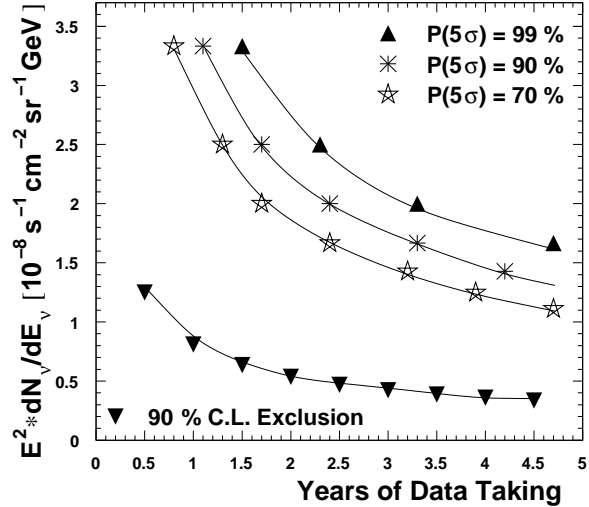,width=.55\textwidth}
\caption[1]{
{\bf Sensitivity to diffuse neutrino fluxes:} Improvement with time.
\label{diffuse_improve_time}
}
\vskip 1.cm
\end{figure}

The sensitivity  obtained after one year of data taking
is already well below the  diffuse bound calculated by  Waxman and 
Bahcall \cite{W&B}.
(This limit holds  
 for optically thin CR sources, under the assumption 
that these sources produce the observed flux of high energy CR.)
A flux at the level $E_{\nu}^2 \times dN_{\nu}/dE_{\nu} =  
2.4 \times 10^{-8}$\,\diffunit~
is needed for a 5-sigma observation after a period of one year. 
This flux is 40 times below the present best established 90\% c.l. upper
limit \cite{diffuse}. 
The improvement with time of the exclusion  and discovery potential of the
detector is
summarized in table \ref{diffuse_sum}.  As the exposure time increases, 
the optimal 
multiplicity cut becomes tighter, resulting in a better separation of 
signal and background. After data is taken over five years, the 
sensitivity is improved by a factor of about 2.5. The 5-sigma detection level
given in table  \ref{diffuse_sum} corresponds to the flux, 
for which the event rate from signal plus background  
exceeds the
5-sigma threshold. The signal strength at which the 5-sigma excess is produced
at a fixed probability, is shown in 
figure \ref{diffuse_improve_time} 
as a function of  time. 
A signal of 
$E_{\nu}^2 \times dE_{\nu}/dN_{\nu} = 1 \times 10^{-8}$\,\diffunit, 
for instance, 
would be detected with a probability of 70\,\% after five years of datataking.

\begin{table}
\renewcommand{\arraystretch}{1.2}
\caption{{\bf Sensitivity to diffuse neutrino fluxes}. 
Expected limits and minimal detectable fluxes  
in units of \diffunit~for a  generic
 $E^{-2}$ source spectrum. Event numbers correspond to a hypothetical source strength of
 $E_{\nu}^{2} \times dN_{\nu}/dE_{\nu} = 1 \times 10^{-7}$\,\diffunit.}
\label{diffuse_sum}
  \begin{center}
   ~\\*[0.2cm]
    \begin{tabular}{lcccccc}
      \hline
      {\bf years} & {\bf $N_{\mathrm{ch}}$ Cut} & {\bf $\langle n_{s} \rangle$} & {\bf $\langle n_{b} \rangle$} & {\bf $\bar{\mu}_{90}$} & {\bf $E^{2} \frac{dN}{d E}({\mathrm{90\%c.l.}})$} & {\bf $E^{2}\frac{dN}{d E}({5\sigma})$}  \\
      \hline \hline
       1            & 227 & 76.4 & 8.0 & 6.1 & $8.1\cdot 10^{-9}$ & $2.6 \cdot 10^{-8}$\\
       3            & 244 & 204.8 & 18.4 & 8.7 & $4.2\cdot 10^{-9}$ & $1.2 \cdot 10^{-8}$\\
       5            & 276 & 272.5 & 18.0 & 8.6 & $3.2 \cdot10^{-9}$ & $9.9 \cdot 10^{-9}$\\
       \hline
\end{tabular}
\\*[1.cm]
\end{center}
\end{table}
\begin{figure}[htp]
\begin{center}
\epsfig{file=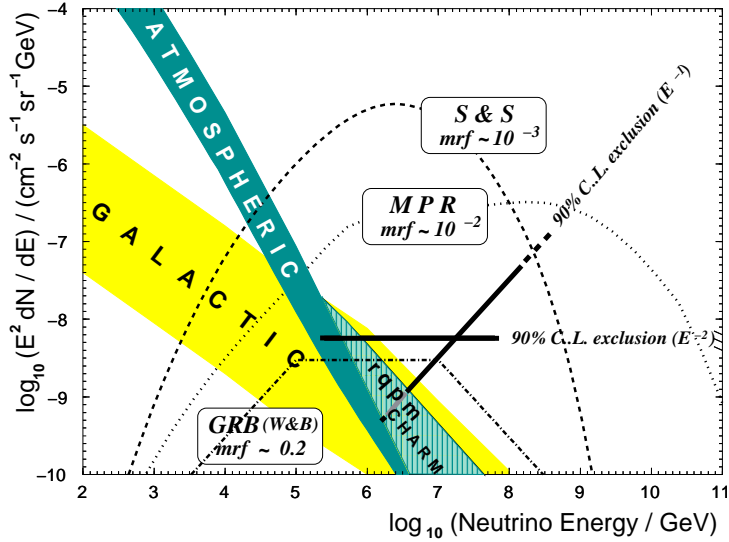,width=.7\textwidth
}
\end{center}
\caption[1]{{\bf Expected sensitivity of the IceCube detector}. Solid lines
indicate the 90\%\,c.l. limit  for various
differential spectra, calculated for a data taking period of three years. 
The lines 
extend over the energy range containing 90\% of the expected signal.
The dashed line indicates the
 Stecker and Salamon model for photo-hadronic interactions in 
AGN cores \protect\cite{SS96}.
The dotted line corresponds to the Mannheim, Protheroe and Rachen  
upper bound on the neutrino emission from photo-hadronic interactions
in AGN jets \protect\cite{MPR98}.
Also shown is the  
GRB model by Waxman and Bahcall \protect\cite{W&B} (dash-dotted line). The 
{\it mrf} was calculated exemplarily for an observation of 500 bursts 
(see section \ref{sec:grb}).
\label{fig:intensities} 
}
\vskip 1.cm
\end{figure}

\begin{table}
\renewcommand{\arraystretch}{1.2}
\caption{{\bf Sensitivity to diffuse neutrino fluxes of various shapes}.
 Expected limits and minimum detectable fluxes  in units of \diffunit~ 
 for different signal hypotheses. Numbers correspond to an exposure time of three years.}
\label{tab:diff.slopes}
  \begin{center}
~\\*[0.2cm]
    \begin{tabular}{lccc}
      \hline
      {\bf Source Model} & {\bf $N_{\mathrm{ch}}$ Cut} & {\bf $\bar{\mu}_{90}$} & {\bf Flux Limit ($ =\Phi \times mrf$)}   \\
      \hline \hline
       $E^{-1}$     & 427 & 3.3 & $dN_{\nu}/dE_{\nu} = 3.1 \cdot 10^{-16} ~(E/\mathrm{GeV})^{-1}$ \\
       $E^{-1.5}$   & 336 & 4.9 & $dN_{\nu}/dE_{\nu} = 1.5 \cdot 10^{-12} ~(E/\mathrm{GeV})^{-1.5}$ \\
       \hline
       \hline
       SS96         & 250 & 8.3 & $\Phi_{\mathrm{SS96}} \times 2.3 \cdot 10^{-3}$ \\
       MPR          & 324 & 5.2 & $\Phi_{\mathrm{MPR}} \times 1.9 \cdot 10^{-2}$ \\
       \hline
\end{tabular}
\\*[1.cm]
\end{center}
\end{table}

Apart from the generic case of an $E^{-2}$ spectrum, which is typical for 
scenarios that involve meson production in interactions of shock  accelerated 
CR with matter,
we have varied the signal slope towards flatter spectra. Such spectra 
would be expected from environments where CR predominantly interact
on photon fields, {\it e.g.} AGN jets \cite{MPR98}. 
The average differential limits for an assumed observation period 
of three years are listed in table \ref{tab:diff.slopes}.

Mannheim, Protheroe and Rachen have calculated an upper bound on the diffuse
neutrino flux arising from photo-hadronic interactions in unresolved
AGN jets in the universe. Their flux bound is shown in figure 
\ref{fig:intensities} labeled {\bf \emph{MPR}}. 
We have determined the model rejection potential for this particular shape 
(and for energies below $10^{8}$\,GeV) to be {\it mrf} $ = 1.9 \cdot 10^{-2}$, 
meaning that \ic~ will be sensitive to fluxes of similar shape, but 
fifty times smaller than the {\bf \emph{MPR}} maximum model.

Finally, we have selected one particular model by Stecker and Salamon 
\cite{SS96} for neutrinos from
 proton interactions on the UV thermal photon field in AGN cores.
The corresponding diffuse flux prediction is labeled {\bf \emph S\&\emph S}
in figure \ref{fig:intensities}. The model rejection factor in this case is 
{\it mrf} $= 2.3 \cdot 10^{-3}$.
Figure \ref{fig:intensities} also shows the 90\% c.l. expected limits 
on an $E^{-2}$ ($E^{-1}$) neutrino flux.

%%%%%%%%%%%%%%%%%%%%%%%%%%%%%%%%%%%%%%%%%%%%%%%%%%%%%%%%%%%%%%%%%%%%%%%%%%%%%%%%%%%%%%%%%%%%%%%%%%%%%%%%%%%%%%%%%%%%%%%%%%%%%%%%%%%%%%%%%%%%%%%%%%%%%%%%%%%%%%%
\subsection{Sensitivity to Point Sources}
\label{sec:point.sens}

An excess of events  from a particular direction in the sky suggests the 
existence of a point source. The ability of the detector to reconstruct 
muon tracks to within 1$^\circ$ of their true direction
 allows a search window to be used that greatly reduces the background, while
retaining a large fraction of the signal. This allows 
for a loosening of the energy separation cut. 

We restrict this analysis to the case of a point source search for
candidate sources in the northern sky. That is, we do not
simulate a cluster or grid search, but we consider the case where 
an angular search bin is fixed by the direction of the candidate 
source under test.
In reality the sensitivity will depend on the declination of the source 
location. For simplicity of presentation 
we calculate averaged event rates 
for all declinations throughout the northern sky.  

We use an  angular search cone of
$1^\circ$ centered about the direction of a hypothetical 
point source.
After application of 
the standard cut selection, we 
optimize the  $N_{\mathrm{ch}}$-cut with 
respect to the model rejection potential for a point 
source following an $E^{-2}$ spectrum.
A cut
at a channel multiplicity of \Nch=30, combined with the angle cut of one 
degree,
leads to the best average flux upper limit of 
$E_{\nu}^2 \times  dN_{\nu}/dE_{\nu} = 5.5 \times 10^{-9}$\,\pointunit~
after one year of data taking. A flux three times greater, will on average, 
produce a 5-sigma signal.

\begin{figure}[htp]
\centering
\epsfig{file=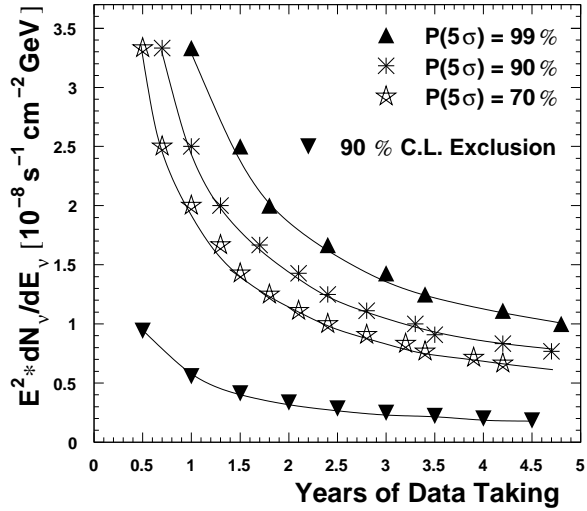,width=.55\textwidth}
\caption[1]{{\bf Sensitivity to point-like neutrino emission:} 
Improvement with time.
\label{point_improve_time}
}
\vskip 1.cm
\end{figure}

Table \ref{pstab} and figure
\ref{point_improve_time}
summarize the improvement of the limit with increased exposure time.
After three years of operation IceCube can be expected to place 
flux limits on potential sources at a level  
$E^{2}_{\nu} \times dN_{\nu}/dE_{\nu} \sim 2.4 \times 10^{-9}$\,\pointunit,
while the discovery probability for a flux three times stronger 
is lager than 70\%. 
After five years of operation a source  emitting a flux of
$E^{2}_{\nu} \times dN_{\nu}/dE_{\nu} \sim 6 \times 10^{-9}$\,\pointunit~would
be observed at 5-sigma significance with a probability of 70\,\%.
~\\

\begin{table}
\renewcommand{\arraystretch}{1.2}
\caption{{\bf Sensitivity to point sources}. 
 Expected limits and minimal detectable fluxes  in units of \pointunit~~ 
 for a  generic
 $E^{-2}$ source spectrum and different exposure times. 
 signal event rates correspond to a hypothetical source
 strength of
 $E_{\nu}^{2} \times dN_{\nu}/dE_{\nu} = 1 \times 10^{-7}$\,\pointunit,
 background event rates include {\tt rqpm} charm neutrinos.}
\label{pstab}
  \begin{center}
~\\*[0.2cm]
    \begin{tabular}{lcccccc}
      \hline
      {\bf years} & {\bf $N_{\mathrm{ch}}$ Cut} & {\bf $\langle n_{s} \rangle$} & {\bf $\langle n_{b} \rangle$} & {\bf $\bar{\mu}_{90}$} & {\bf $E^{2} \frac{dN}{d E}({90\% c.l.})$} & {\bf $E^{2}\frac{dN}{d E}({5\sigma})$}  \\
      \hline \hline
       1 & 30 & 62.8 & 1.4 & 3.6  & $5.5 \cdot 10^{-9}$ & $1.7 \cdot 10^{-8}$\\
       3 & 40 & 142.3 & 1.3 & 3.5  & $2.4 \cdot 10^{-9}$ & $7.2 \cdot 10^{-9}$\\
       5 & 42 & 213.7 & 1.4 & 3.6  & $1.7 \cdot10^{-9}$ & $4.9 \cdot 10^{-9}$\\
       \hline
\end{tabular}
\\*[1.cm]
\end{center}
\end{table}

\begin{figure}[htp]
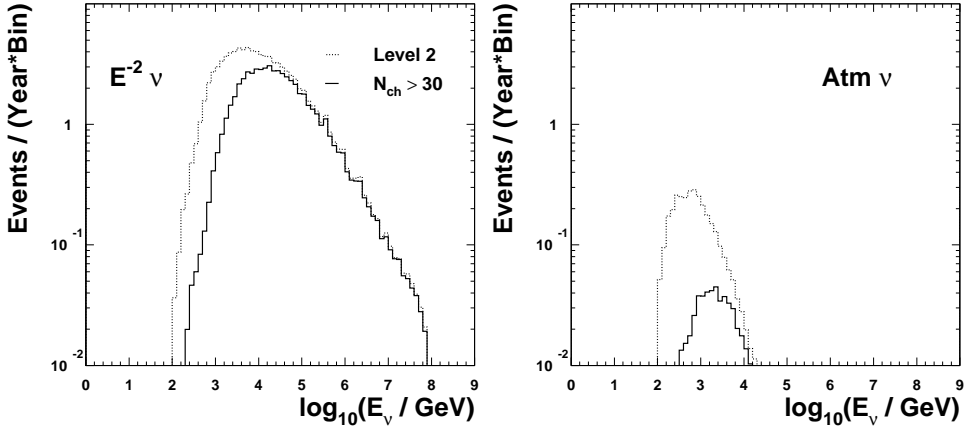

\begin{center}
\epsfig{file=./plots/nupass.trueen.point.1logy.fin.selec.FETTBW.epsi,
width=.45\textwidth
}
\hskip .05cm
\epsfig{file=./plots/nupass.trueen.point.0logy.fin.selec.FETTBW.epsi,
width=.45\textwidth
}
\end{center}
\caption{{\bf Energy spectra of selected neutrinos} for a $E^{-2}$ 
source ({\it left})
 and atmospheric neutrinos including {\tt rqpm} charm
({\it right}). The selection corresponds to 
level\,2 cuts (dotted lines) and additional application of 
the optimized cut \Nch$>30$ 
(full lines).
\label{fig:point.energy.spectra}
}
\vskip 1.cm
\end{figure}

Figure \ref{fig:point.energy.spectra} shows the energy spectra of both the 
remaining signal events and the remaining events from the atmospheric neutrino 
background
after applying standard cuts and after cutting at $N_{\mathrm{ch}} > 30$.
This cut results in an effective energy threshold of 
1\,TeV.
Since most of the signal is in the TeV region, the  
energy cut-off of the Monte Carlo simulation 
has negligible impact on the result. 
Above results are valid for the {\tt rqpm} prediction for prompt neutrinos. 
Using the {\tt TIG} model improves the sensitivity by about 2\,\%.

As for the diffuse signal, we have tested different exponential slopes for
the signal hypothesis. The results listed in table 
\ref{tab:point_slopes} correspond to three years of data taking.
     
\begin{table}
\renewcommand{\arraystretch}{1.2}
  \begin{center}
   \caption{{\bf Sensitivity to point source fluxes of various exponential slopes}.
Expected limits and minimal detectable fluxes  in units of \pointunit~ 
for different signal hypotheses. Numbers correspond to an exposure time of three years.}
\label{tab:point_slopes}
~\\*[0.2cm]
    \begin{tabular}{lccl}
      \hline
      {\bf Source Model} & {\bf $N_{\mathrm{ch}}$ Cut} & {\bf $\bar{\mu}_{90}$} & {\bf Flux Limit ($=\Phi \times mrf$)}   \\
      \hline \hline
       $E^{-1}$     & 58 & 2.7 & $dN_{\nu}/dE_{\nu} = 2.4 \cdot 10^{-15} ~(E/\mathrm{GeV})^{-1}$ \\
       $E^{-1.5}$   & 49 & 2.9 & $dN_{\nu}/dE_{\nu} = 4.5 \cdot 10^{-12} ~(E/\mathrm{GeV})^{-1.5}$ \\
       $E^{-2}$     & 40 & 3.5 & $dN_{\nu}/dE_{\nu} = 2.4 \cdot 10^{-9} ~(E/\mathrm{GeV})^{-2}$ \\
       $E^{-2.5}$   & 24 & 6.1 & $dN_{\nu}/dE_{\nu} = 3.8 \cdot 10^{-5} ~(E/\mathrm{GeV})^{-2.5}$ \\
       \hline
\end{tabular}
\\*[1.cm]
\end{center}
\end{table}

\subsection{Gamma Ray Burst Sensitivity}
\label{sec:grb}

Although the progenitors of GRBs are unknown, observations indicate 
the existence of a fireball. The coexistence of nucleons and 
photons in the fireball may result in the production of neutrinos.

 Waxman and Bahcall \cite{W&B}
calculated the expected flux of neutrinos
 from the sum of all GRBs by assuming that they are the source of
the observed flux of cosmic rays. 
The Waxman-Bahcall model 
results in a broken power-law  neutrino spectrum given by
\begin{equation}
       \frac{dN_{\nu}}{dE_{\nu}} = 
         \left \{
           \begin{array}{lccr} 
           \frac{A}{E_{\nu} E_{\nu}^{\mathrm{b1}}} & , &   E_{\nu} < E_{\nu}^{\mathrm{b1}} \\
           \frac{A}{E_{\nu}^{2}} & , &  E_{\nu}^{\mathrm{b1}} <  E_{\nu} < E_{\nu}^{\mathrm{b2}}, \\
           \end{array}
           \right.\\
\label{eq:wbgrb}
\end{equation}
where the break energy $E_{\nu}^{\mathrm{b1}}$ lies at $\sim\,10^{5}$\,GeV. 
Above $E_{\nu}^{\mathrm{b2}} = 10^{7}$\,GeV the spectrum steepens again by one power in energy.
With  a full sky GRB rate of $\sim$1000 per year, as assumed by Waxman and Bahcall, 
the normalization constant in equation \ref{eq:wbgrb} would amount to
$A \sim 3 \times 10^{-9}$\,\diffunit.\footnote{A more recent calculation 
yielded a normalization constant which is about three times larger 
\cite{waxmangrb}.}
This neutrino flux is shown
in figure \ref{fig:intensities} labeled {\bf \emph{GRB}}. 
It appears to be below the diffuse flux sensitivity level of \ic. However,
the search for neutrinos accompanying a GRB is essentially background-free, because one 
looks for neutrino events that are coincident in direction and time to
the satellite observation.  

The search for neutrinos from GRBs involves summing over the observation
time and spatial search windows for many separate bursts.
 For this analysis we
have used a hypothetical observation duration of 10 seconds and 
a spatial search cone of $10^\circ$  centered about the 
direction of each GRB.
We have only considered events in the
northern sky, where we can be sure that the search will not be limited by 
downgoing CR muon background.
From 500 bursts 
in $2\pi$\,sr we would expect 13 neutrino-induced up-going 
muons per year after applying standard level\,2 quality cuts 
(table \ref{tab:cuts}).
The background of atmospheric neutrinos is strongly reduced by
the spatial and temporal coincidence requirements. With almost full retention 
of the signal,  the atmospheric neutrino background expectation
is reduced to roughly 0.1 event. 
The smallness of the background expectation allows one to exclude 
signals of mean intensity $\langle n_s \rangle = \bar{\mu}_{90}$ = 2.5 
events per year at 90\% classical
confidence, meaning that the experiment will be sensitive to a neutrino flux roughly
five times smaller than the flux calculated by Waxman and Bahcall (equation 
\ref{eq:wbgrb}). This also means that the observation of 100 bursts 
would suffice to exclude the Waxman and Bahcall model.
A 5-sigma detection would require the observation of $n_{0} = 5$ events,
which corresponds to the mean number of events expected from 203 bursts. 
In this case the probability to actually observe a 5-sigma excess is about 
58\%. With  500 bursts this probability climbs to  99\%.
The time period after which we can expect a detection depends on the
efficiency of the gamma ray observations, 
since the search strategy requires the GRBs to be triggered by satellites. 
Assuming that future gamma ray observations will provide a few hundred triggered burst 
per year, we can conclude 
that IceCube has
excellent prospects
to reveal the neutrino signal possibly emerging from  GRBs 
within a very short time: The analysis of data taken over one year 
would presumably suffice to yield a 5-sigma signal
(provided the model by Waxman and Bahcall predicts neutrino 
fluxes at the right scale). Moreover, the sensitivity given above is 
obtained when employing the most conservative search strategy, 
namely searching only one hemisphere for the signal of 
up-going neutrinos. However,
one can be convinced that the drastic background reduction   
due to the small observation time window will result in a 
sizable acceptance also for downward signal.

\subsection{Systematic Uncertainties and Possible Improvements }
The present systematic uncertainty of the given  flux limits
is dominated by three components. The largest is the
uncertainty in the angular dependence of the OM sensitivity,
including the effect of the refrozen ice around the OM.
A local increase in light scattering from air bubbles trapped
in the vicinity of the OM translates into a modulation
of its angle-dependent acceptance. This component is followed
in size by uncertainties in the absolute OM sensitivity
and in the optical properties of the bulk ice.
For the comparatively small AMANDA-B10 array the inclusion
of all components of uncertainty weakens the point source
flux limit by 25\% compared to standard simulation values \cite{steve}. 
The variation of some of these
parameters in simulations of the larger AMANDA-II array
and for IceCube indicates that for larger arrays the systematic
uncertainties of the basic input parameters become less important,
with the exception of muon energies close to
the detection threshold. For instance, increasing
the absolute OM sensitivity in IceCube by a factor of 2 results
in a  25 (10) \% larger effective area at 1 (10) TeV.
Taking into account that uncertainties in limits depend
more weakly than linearly on uncertainties in effective area
\cite{steve}, we estimate the overall
uncertainties of the $E^{-2}$ limits derived above
to be at most 20\%.

On the other hand, we may expect that improved detector
properties for IceCube as compared to AMANDA will result
in smaller systematic uncertainties and a better performance. 
First,
the use of glass spheres and PMT glass with better UV transparency or,
alternatively, a covering of the glass spheres with wavelength shifter,
will enhance the OM sensitivity in the UV region and result in a better
light collection. This will particularly increase the sensitivity and
angular resolution at {\it low energies}. 
Information obtained from the full waveform will improve both the
angular resolution and the energy reconstruction at {\it high energies}.
Waveform information will be used in AMANDA from 2003 on,
and methods to make efficient use of the corresponding information are
under development. Finally, the information from the
IceTop surface array will enhance the rejection power with respect
to downward moving atmospheric muons. This method, unique to
IceCube, is expected to be
particularly helpful for muons from coincident, independent air showers
and would allow a loosening other rejection criteria, thereby
enhancing the signal efficiency.

\section{Summary}
We have described the performance
of the IceCube detector in searching for muons
from extraterrestrial neutrinos in the TeV-PeV
energy range.

A Monte Carlo simulation of a realistic model detector was used to 
assess the sensitivity of the experiment.
We have simulated both neutrino-induced muons and muons produced from 
cosmic-ray interactions in the atmosphere with sufficient statistics to 
establish event selection criteria and infer event rates to be expected from 
each event class. The trigger rate due to down-going muons produced in the
atmosphere 
was found to be 1.7\,kHz including a 50\,Hz rate due to non-correlated 
air-showers that produce time-coincident muons within the detector. 
Muons induced by atmospheric neutrinos are expected to cause 
about 0.8 million  triggers per year.
A benchmark flux of 
$E_{\nu}^2 \times dN_{\nu}/dE_{\nu}=10^{-7}$\,\diffunit~for the 
diffuse signal of astrophysical neutrinos results in 3300 triggers per year.
Roughly a third of them pass quality cuts which at the same time
reduce the background rate
from misreconstructed downward muon tracks to the level of well
reconstructed upward muons from atmospheric neutrinos. 

In order to quantify the detector acceptance, we have computed the effective
detector area. After applying a set of 
standard quality criteria, the effective area 
exceeds one square kilometer for upward-going
muons with energies of 10\,TeV and above. At this stage, 50\,\% of all 
muons of this energy will be reconstructed with an accuracy of 
0.8$^\circ$ or better. For energies above 100 TeV, the
angular acceptance with respect to well identified
extraterrestrial neutrinos extends  above the 
horizon and the effective area 
reaches 0.6 km$^2$  for downgoing muons in the PeV range.
This means that at high energies IceCube can observe a 
large part of the Galaxy, including the galactic center.

In order to quantify the sensitivity to fluxes of astrophysical 
neutrinos, we have determined the flux normalization for a 
generic $E^{-2}$ differential energy spectrum that correspond to a detection 
with  
5-sigma significance, or, in absence of signal, a 90\,\%\, c.l. 
limit. 
We found a diffuse source strength of $\dNdE = 10^{-8}$\,\diffunit~ for
the 5-sigma detection level and 
$4 \times 10^{-9}$\,\diffunit~ for the exclusion potential of the 
detector, given an observation time of three years.
This is two orders of magnitude below present experimental limits.
For point-like neutrino emission we found that, after three years, a flux of 
$\dNdE = 7 \times 10^{-9}$\,\pointunit~ would result in a 5-sigma excess over
background, while a flux of $\dNdE = 2 \times 10^{-9}$\,\pointunit~ can be
excluded at 90\,\%\,c.l.
Both numbers are averaged over all declinations throughout the northern sky.
Integrated over all neutrino energies above 1 TeV, these fluxes
transform to
$F_{\nu}$($>$\,1\,TeV) = $ 7 (2) \times 10^{-12}$ cm$^{-2}$\,s$^{-1}$.

We have also calculated the potential of IceCube to detect neutrinos
in coincidence with gamma ray bursts, following the model
of Waxman and Bahcall. We found that a 5-sigma 
signal is expected from the observation of about 
200 bursts, while an observation 
of 100 bursts would suffice to rule out the Waxman and Bahcall model.

\ack{
This research was supported by the following agencies: National
Science Foundation--Office of Polar Programs, National Science
Foundation--Physics Division, University of Wisconsin Alumni Research
Foundation, USA; Swedish Research Council, Swedish Polar Research
Secretariat, Knut and Alice Wallenberg Foundation, Sweden; German
Ministry for Education and Research, Deutsche Forschungsgemeinschaft
(DFG), Germany; Fund for Scientific Research (FNRS-FWO), Flanders
Institute to encourage scientific and technological research in
industry (IWT), Belgian Federal Office for Scientific, Technical
and Cultural affairs (OSTC), Belgium; Inamori Sceience Foundation,
Japan; FPVI, Venezuela; The Netherlands Organization for
Scientific Research (NWO).
}

\end{document}